# JTAG-based Remote Configuration of FPGAs over Optical Fibers

Binwei Deng,[a,b] Chonghan Liu,[b] Jinghong Chen,[c,d] Kai Chen,[e] Datao Gong,[b] Di Guo,[f,b] Suen Hou,[g] Deping Huang,[c,d] Xiaoting Li,[h,b] Tiankuan Liu[b,*] Ping-Kun Teng,[g] Annie C. Xiang,[b] Hao Xu,[e] Yang You,[c] Jingbo Ye[b]

[a] *School of Electric and Electronic Information Engineering, Hubei Polytechnic University,*
  *Huangshi, Hubei 435003, P. R. China*

[b] *Department of Physics, Southern Methodist University,*
  *Dallas, TX 75275, USA*

[c] *Department of Electrical Engineering, Southern Methodist University,*
  *Dallas, TX 75275, USA*

[d] *Department of Electrical Engineering, University of Houston*
  *Houston, Texas, 77004*

[e] *Department of Physics, Brookhaven National Laboratory*
  *Upton, NY 11973*

[f] *Department of Modern Physics, University of Science and Technology of China,*
  *Hefei, Anhui 230026, P. R. China*

[g] *Institute of Physics, Institute of Physics, Academia Sinica,*
  *Nangang 11529, Taipei, Taiwan*

[h] *Department of Physics, Central China Normal University,*
  *Wuhan, Hubei 430079, P.R. China*
  *E-mail:* tliu@mail.smu.edu

ABSTRACT: In this paper, a remote FPGA-configuration method based on JTAG extension over optical fibers is presented. The method takes advantage of commercial components and ready-to-use software such as iMPACT and does not require any hardware or software development. The method combines the advantages of the slow remote JTAG configuration and the fast local flash memory configuration. The method has been verified successfully and used in the Demonstrator of Liquid-Argon Trigger Digitization Board (LTDB) for the ATLAS liquid argon calorimeter Phase-I trigger upgrade. All components on the FPGA side are verified to meet the radiation tolerance requirements.

# Contents

## 1. Introduction

The ATLAS Liquid Argon Calorimeter (LAr) Phase-1 trigger upgrade [1] has been proposed to enhance the physics reach of the ATLAS experiment. A LAr Trigger Digitizer Board (LTDB) is being developed to read out up to 320 detector channels and transmit all the data from the front-end detector to the back-end counting room through 40 optical fibers. The LTDB, once installed, will operate in a harsh radiation environment [2]. The LTDB will be implemented in radiation-tolerant components [3-5] such as ASICs. An LTDB prototype, called the Demonstrator, based on gigabit-transceiver-embedded FPGAs (Xilinx Kintex 7 series, part number XC7K325TFFG900) has been developed to evaluate the LTDB functions. The FPGAs need to be configured after each power cycle or after a single-event functional interrupts (SEFI) occurs. The distance between the Demonstrator and the counting room is about 70 meters. It is indispensable to develop a remote configuration method.

    The JTAG mode [6] is one of the most commonly used configuration modes. In addition to the configuration of FPGAs, the JTAG mode provides extra capabilities unavailable in other configuration modes. These extra capabilities include the configuration of flash memories with serial peripheral bus (SPI) interface through an FPGA and the logic analyzer tool called ChipScope Pro. Therefore, we choose the JTAG mode to configure the FPGA.

    The simplest way to extend JTAG signals is to use one fiber for each JTAG signal. Since the JTAG interface has at least four signals (TCK, TMS, TDI and TDO), at least four optical fibers are needed. Moreover, JTAG signals are not DC balanced and not suitable for modern commercial optical transceiver modules which are usually internally AC coupled [7-8].

    A commonly used approach to configure remote FPGAs is to develop a local JTAG master and transmit the configuration contents over optical fibers [9-11] to the local JTAG master. This approach can configure FPGAs. However, it is not easy to develop a JTAG master, let alone a JTAG master that needs to operate in a radiation environment.

    In this paper, a remote FPGA configuration approach based on JTAG extension over optical fibers has been proposed and verified. The approach has been used in the LTDB Demonstrator. The design of the remote configuration system for the LTDB Demonstrator and the radiation qualification of related components are presented.

## 2. The proposed JTAG extension approach

The remote configuration approach we propose is shown in Figure 1. Three JTAG signals (TMS, TCK, and TDI) coming out of the Xilinx download cable (Xilinx Platform Cable USB II) are encoded and serialized into a high-speed serial data signal. The encoder and the serializer are integrated into a transmitter which has multiple parallel input signals (TxD0, TxD1, TxD2, …) and a high-speed serial data output signal. TMS, TCK, and TDI are connected to three parallel input signals of the transmitter. The high-speed serial data signal is converted to the optical signal in an optical transmitter (E/O) module and transmitted through an optical fiber. The optical signal is converted into the high-speed serial data electrical signal in an optical receiver (O/E) module. The high-speed serial data signal is deserialized and decoded to recover the corresponding JTAG signals before they are connected to the FPGA. The deserializer and the decoder are integrated into an receiver integrated circuit which has a high-speed serial data signal and multiple parallel outputs (RxD0, RxD1, RxD2, …). The other direction operates in the same way except that only TDO is transmitted. We only need two fibers, each for one direction.

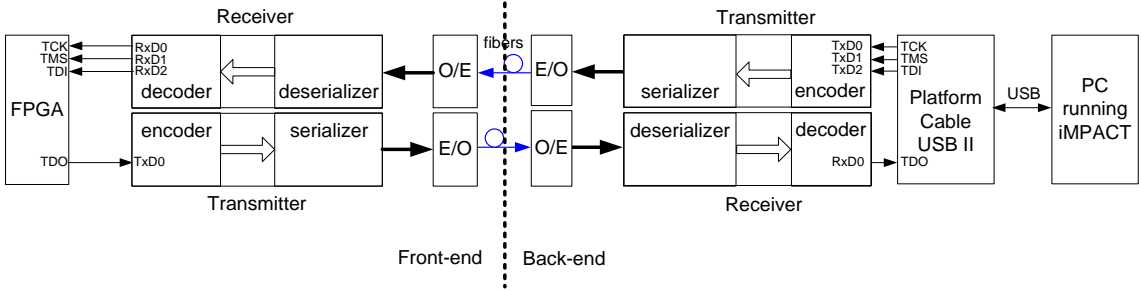

Figure 1: the block diagram of the proposed remote configuration approach

Due to the propagation delay induced by long fibers and transceivers, the frequency of TCK may need to be slower than that without any JTAG extension. According to IEEE standard 1149.1, on the FPGA side, TDO changes on the falling edge of TCK, while TMS and TDI are sampled on the rising edge of TCK. On the JTAG Xilinx download cable side, TMS and TDI change on the falling edge of TCK, while TDO is sampled on the falling edge of TCK [12]. In order for TSM/TDI/TDO to have efficient setup time, the propagation delay induced by the transceiver and optical fibers must meet the following relationships:

$$\frac{1}{2f_{TCK}} \geq T_{CQcable(max)} + T_{SUfpga(min)}, \quad (1)$$

$$\frac{1}{f_{TCK}} \geq T_{CQfpga(max)} + 2T_{PD} + T_{SUcable(min)}. \quad (2)$$

Where $f_{TCK}$ is the frequency of TCK, $T_{PD}$ is the propagation delay of the whole link, including the transmitter, the optical transmitter, the fiber, the optical receiver, the receiver, and relevant traces on printed circuit boards, $T_{CQcable(max)}$ and $T_{CQfpga(max)}$ are the maximum clock-to-output time ($T_{CQ}$) of the output D flip-flop (DFF) at the download cable side and on the FPGA side, respectively; $T_{SUcable(min)}$ and $T_{SUfpga(min)}$ are the minimum setup time ($T_{SU}$) of the input DFF at download cable side and on the FPGA side, respectively. $T_{CQ}$ and $T_{SU}$ can be found in the download cable and the FPGA data sheets. The sum of $T_{CQ}$, $T_{SU}$, and the propagation delay of optical transceiver and the PCB traces is estimated to be no greater than 20 ns. The latency of the TLK2501 transceiver is no greater than 58 ns [13]. The maximum fiber lengths at different configuration clock frequencies are estimated and listed in Table 1. A latency of 5 ns per meter for the optical fiber is assumed in the table.

GBTX [14], a radiation-tolerant 4.8-Gbps transceiver, may be used to replace TLK2501. Since GBTX does not support the LVCMOS format which all JTAG signals belong to, level translators (for example, part numbers SN75LVDS386 and SN75LVDS387 produced by Texas Instruments) are necessary for the interface between the Kintex-7 FPGA and GBTX. The latency of GBTX (including the corresponding GBT FPGA used on the back end and the level translators) is measured to be about 292 ns.

When TLK2501 is used on both the front end and the back end with 70-m duplex fibers, 750 KHz can be chosen. For applications which require longer fibers and transceivers with longer latency, Digilent USB JTAG cable (part number 250-003P produced by Digilent Inc.), which supports a minimum clock frequency of 250 KHz, can be used. The frequency of TCK is chosen in the cable communication setup of iMPACT.

Table 1: The maximum fiber length at different configuration clock frequency

| frequency (MHz) | | 6 | 3 | 1.5 | 0.75 | 0.5 | 0.25 |
|---|---|---|---|---|---|---|---|
| Maximum fiber length (m) | TLK2501 | 3 | 20 | 53 | 120 | 186 | 386 |
| | GBTX | | | 6 | 73 | 140 | 340 |

The configuration approach has been verified with two commercial SFP+ optical transceivers modules, two Evaluation Modules (EVMs) of the transceivers TLK2501 (each of which includes an 8B/10B encoder, an 8B/10B decoder, a serializer, and a deserializer), 100-meter duplex multimode fibers, and a Xilinx Kintex-7 FPGA KC705 Evaluation Module (EVM) [15]. A picture of the verification test setup is shown in figure 2.

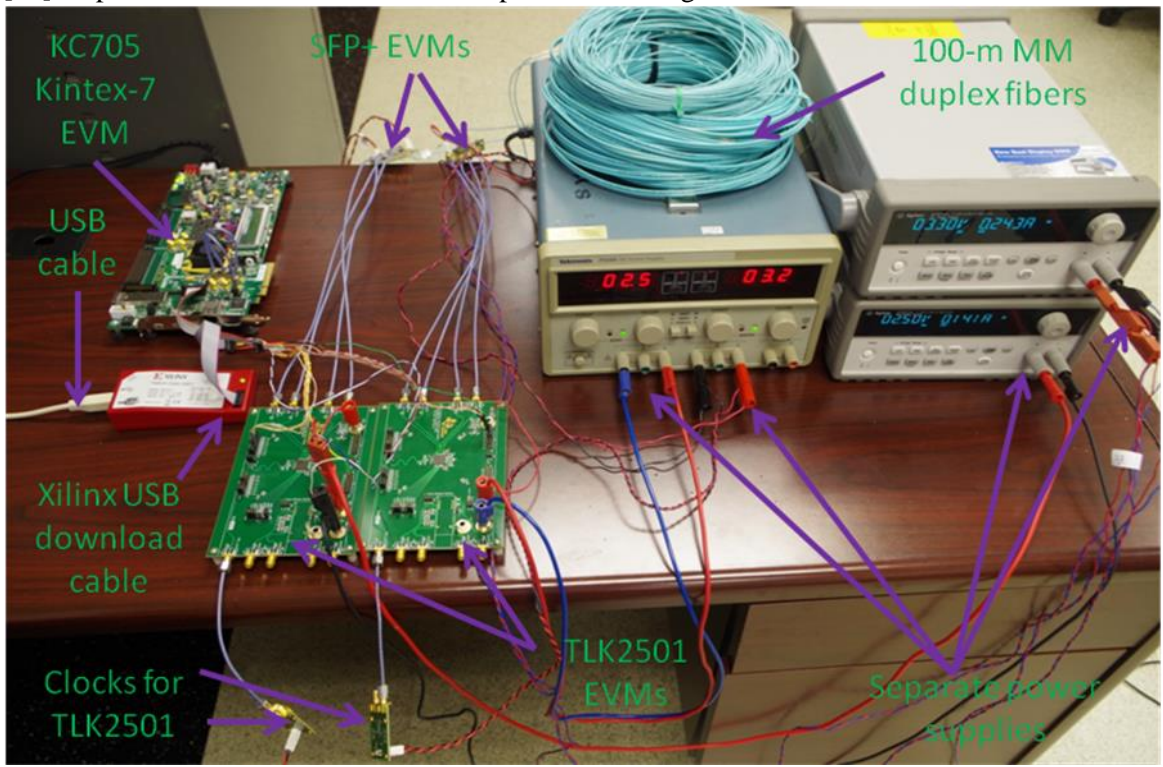

Figure 2: the picture of the verification test setup

The proposed remote configuration approach has been used in the LTDB Demonstrator. The block diagram of the remote FPGA configuration system is shown in figure 3. Two transceivers TLK2501 are used on each LTDB Demonstrator, whereas the 8B/10B encoder/decoder and the transceiver are implemented in the Virtex6 FPGA of a Gigabit Link Interface Board (GLIB) [16] on the back end. Two Xilinx Kintex-7 series FPGAs (part number XC7K325TFFG900) are configured through a single pair of fibers and a TLK2501. A quad-channel optical transceiver module (part number AFBR-79EIDZ produced by Avago Technologies) [8] is used on the LTDB Demonstrator. On the back end, a Digilent USB JTAG cable is connected to the Virtex-6 FPGA through an FMC extension board. A 1:4 switch is implemented in the Virtex FPGA on the GLIB to select which FPGA on the front end to be configured. Two SFP+ optical transceivers are used on the back end. As a backup solution, we use a flash memory with the Serial Peripheral Interface (SPI) for each FPGA on the Demonstrator. The FPGA can be configured from the flash memory locally and quickly. The flash memory can also be remotely configured via the FPGAs and the fibers.

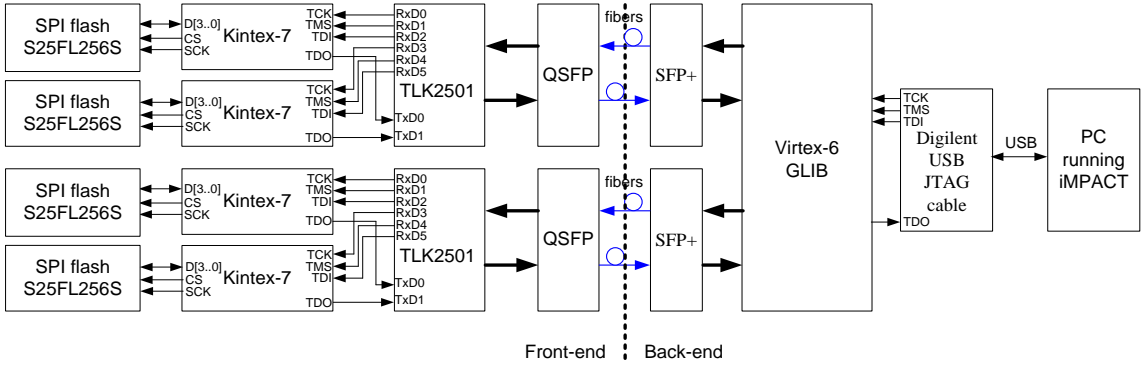

Figure 3: the block diagram of the remote configuration on the LTDB Demonstrator

Our method combines the advantages of slow remote configuration and fast local configuration. The method takes advantage of commercial components and ready-to-use software such as iMPACT and does not require any hardware or software development.

## 3. Radiation Tolerance Qualification

The Demonstrator will operate at a low luminosity of 33.3 fb$^{-1}$/year for one to three years (2015-2017). All components in use for the Demonstrator must tolerate the radiation level for at least one year but three years is preferable. The radiation tolerance criteria include Total Ionizing Dose (TID), Non-Ionizing Energy Loss (NIEL), and Single-Event Effects (SEE). The simulated radiation level [2, 17] at the position where the LTDB Demonstrator is installed, safety factors [18], and radiation tolerance criteria [2] are listed in the table 1.

We have tested a QSFP optical transceiver and a flash memory in X-rays or protons for TID and neutrons or protons for Single-Event Effects (SEE). TLK2501 has been tested in a proton beam before [18]. We have not tested any component for Non-Ionizing Energy Loss (NIEL) (the QSFP was tested in a proton beam, but the fluence was not high enough to qualify the module). Most modern integrated circuits, including TLK2501, the flash memory under test, the microcontroller used inside the QSFP, are fabricated in CMOS technologies which are naturally insensitive to NIEL [2]. The VCSELs and photodiodes used in the QSFP AFBR-79EIDZ may be sensitive to NIEL, but previous tests [20] have shown that commercial VCSELs and photodiodes degrade little at the low fluence of $1.6\times10^{12}$ cm$^{-2}$. Optical fibers have been qualified before [21-22].

Table 1: the simulated radiation level, safety factors, and radiation tolerance criteria

|  | Simulated radiation level | Safety factor | | | Radiation tolerance criteria (3 years) |
|---|---|---|---|---|---|
|  |  | Simulation | Low dose rate effect | Lot-lot variation |  |
| TID | 3.0 Gy(SiO2) | 1.5 | 5 | 4 | 90 Gy(SiO2) |
| NIEL | $2.0\times10^{11}$ cm$^{-2}$ 1-MeV equ. neutrons in Si | 2 | 1 | 4 | $1.6\times10^{12}$ cm$^{-2}$ neutrons |
| SEE | $2.8\times10^{10}$ cm$^{-2}$ >20 MeV hadrons | 2 | 1 | 4 | $2.3\times10^{11}$ cm$^{-2}$ hadrons |

The QSFP optical transceiver AFBR-79EIDZ was tested in a neutron beam at the Los Alamos Neutron Science Center (LANSCE) at Los Alamos National Laboratory. The maximum neutron energy is 800 MeV. The energy spectrum of neutrons is similar to that simulated in the ATLAS application environment [17]. The test setup is shown in figure 4. A Pseudo-Random Binary Sequences (PRBS) (pattern $2^7$-1) at 5 Gbps was generated in a Kintex-7 FPGA. Three channels of data went through the QSFP transmitter. One channel went back to the same FPGA through the QSFP receiver, while the other two channels went to another FPGA through two SFP+ optical receivers. The received data were checked and any errors were recorded in receiver FPGA [23]. The QSFP was put inside the neutron beam, while the other components were outside the beam. During the test of about 43 hours, the total fluence was $4.31\times10^{10}$ cm$^{-2}$. No single event upset (SEU) or SEFI errors were observed during the test. The SEE cross section was estimated to be less than $2.3\times10^{-11}$ cm$^2$/device.

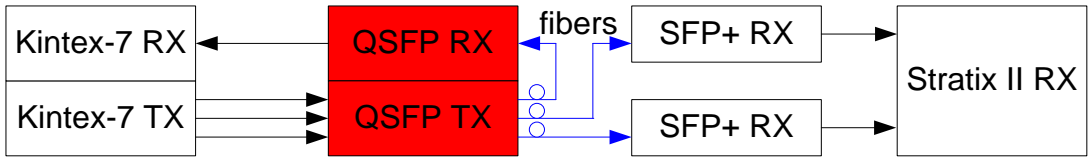

Figure 4: the test setup of QSFP optical transceiver

The QSFP optical transceiver was tested in x-rays to the total dose of 81 Gy(SiO2) with an in-house x-ray biological irradiator (part number X-RAD iR-160 produced by Precision X-ray, Inc.). The maximum x-ray energy is 160 keV. The peak x-ray energy is about 30 keV with a 2-mm aluminum filter. No performance degradation was observed.

The QSFP optical receiver was tested in a 223-MeV proton beam at the Francis H. Burr Proton Therapy Center at of Massachusetts General Hospital (MGH). The test setup was similar to that at LANSCE except that only one receiver channel was tested. The proton energy was 223 MeV. The profile of the proton beam was about 2 cm (diameter) with 90% uniformity. The module was irradiated at $1.3\times10^7$ cm$^{-2}$s$^{-1}$ for 30 minutes and $8.0\times10^7$ cm$^{-2}$s$^{-1}$ for another 11 minutes. No SEU or SEFI errors were observed during the test. The total fluence was $7.5\times10^{10}$ cm$^{-2}$. The SEE cross section was estimated to be less than $1.3\times10^{-11}$ cm$^2$/device. During the test, the TID was accumulated to 42 Gy(SiO2). The accumulated NIEL fluence was $7.1\times10^{10}$ 1-MeV equivalent neutrons cm$^{-2}$.

Two SPI flash memories (part number S25FL256SAGMFIR01 produced by Spansion) were tested in the 223-MeV proton beam at MGH. The first memory was irradiated for one hour to reach the fluence of $9.7 \times 10^{10}$ cm$^{-2}$ and the TID of 63 Gy(SiO2). The second memory was irradiated for 7 minutes to reach the fluence of $3.6 \times 10^{11}$ cm$^{-2}$ and the TID of 182 Gy(SiO2). The memories were powered during the test. The power consumptions of these two memories were monitored during the tests and no significant changes were observed. The contents of the memories before and after irradiation were compared. No errors were observed. The SEE cross section was estimated to be less than $3.3 \times 10^{-12}$ cm$^2$/device.

The TID test results are summarized in Table 2. Two translators which may be used with GBTX are also listed in the table. All components can operate in the ATLAS environment for at least 2.7 years (the QSFP optical transceiver) or longer than three years (all the others).

Table 2: the test results of TID tests

| Device | Part number | Radiation type | TID (Gy) | Degradation | Max Icc change |
|---|---|---|---|---|---|
| QSFP optical transceiver | AFBR-79EIDZ | X-ray | 81 | No | Not measured |
| Transceiver | TLK2501 [19] | X-ray | 1300 | No | +150% |
| SPI flash memory | S25FL256S | Proton | 185 | No | 1% |
| translator | SN75LVDS386 | X-ray | 100 | No | -8.3% |
| translator | SN75LVDS387 | X-ray | 210 | No | +12% |

Table 3: the test results of SEE tests

| Device | Part number | Radiation type | Non-SEFI SEU | | SEFI | |
|---|---|---|---|---|---|---|
| | | | σ (cm$^2$) | Est. err rate (1/yr) | σ (cm$^2$) | Est. err rate (1/yr) |
| QSFP optical transceiver | AFBR-79EIDZ RX only | Proton | <1.3×10$^{-11}$ | <1.0 | <1.3×10$^{-11}$ | <1.0 |
| | AFBR-79EIDZ TX + RX | Neutron | <2.3×10$^{-11}$ | <1.8 | <2.3×10$^{-11}$ | <2.0 |
| Transceiver | TLK2501 [18] | Proton | 1.1×10$^{-11}$ | 0.83 | 8.0×10$^{-12}$ | 0.61 |
| SPI flash memory | S25FL256S | Proton | <3.3×10$^{-12}$ | <0.23 | <3.3×10$^{-12}$ | <0.23 |
| translator | SN75LVDS386 | Neutron | <8.2×10$^{-12}$ | <0.62 | <8.2×10$^{-12}$ | <0.62 |
| translator | SN75LVDS387 | Neutron | <8.1×10$^{-12}$ | <0.62 | <8.1×10$^{-12}$ | <0.62 |

The SEE test results are summarized in Table 3. Two translators which may be used with GBTX are also listed in the table. The error rate of each component in the application operation environment is estimated. The upper limits of cross section and estimated error rate are listed in the table if no error was observed during the test. All components have less than one error per year and meet the radiation tolerate requirements.

## 4. Conclusion

A remote FPGA-configuration method based on JTAG extension over optical fibers is presented. The Demonstrator using the approach has been installed in the ATLAS liquid argon calorimeter upgrade Demonstrator. All components on the FPGA side are verified to meet the radiation tolerance requirements.

## Acknowledgments

This work is supported by US-ATLAS R&D program for the upgrade of the LHC, the US Department of Energy Grant DE-FG02-04ER1299, National Science Council in Taiwan, and Hubei Provincial Natural Science Foundation of China (Grant Number 2014CFC1093). The authors would like to express the deepest appreciation to Ms. Tanya Herrera, Dr. Steve Wender, and Dr. Ron Nelson from the LANSCE of Los Alamos National Laboratory, Dr. Hucheng Chen, Dr. James Kierstead, and Dr. Helio Takai from Brookhaven National Laboratory, Dr. Mike Wirthlin from Young Braham University, Dr. Ethan Casio from Massachusetts General Hospital for beneficial discussions and kind help during the irradiation tests.